# Nonlinear inverse synthesis for high spectral efficiency transmission in optical fibers

Son Thai Le[*], Jaroslaw E. Prilepsky, and Sergei K. Turitsyn

*Aston Institute of Photonic Technologies, Aston University, Birmingham B4 7ET, UK*
*[*let1@aston.ac.uk](mailto:let1@aston.ac.uk)*

**Abstract:** In linear communication channels, spectral components (modes) defined by the Fourier transform of the signal propagate without interactions with each other. In certain nonlinear channels, such as the one modelled by the classical nonlinear Schrödinger equation, there are nonlinear modes (*nonlinear signal spectrum)* that also propagate without interacting with each other and without corresponding nonlinear cross talk, effectively, in a linear manner. Here, we describe in a constructive way how to introduce such nonlinear modes for a given input signal. We investigate the performance of the nonlinear inverse synthesis (NIS) method, in which the information is encoded directly onto the continuous part of the nonlinear signal spectrum. This transmission technique, combined with the appropriate distributed Raman amplification, can provide an effective *eigenvalue division multiplexing* with high spectral efficiency, thanks to highly suppressed channel cross talk. The proposed NIS approach can be integrated with any modulation formats. Here, we demonstrate numerically the feasibility of merging the NIS technique in a burst mode with high spectral efficiency methods, such as orthogonal frequency division multiplexing and Nyquist pulse shaping with advanced modulation formats (e.g., QPSK, 16QAM, and 64QAM), showing a performance improvement up to 4.5 dB, which is comparable to results achievable with multi-step per span digital back propagation.

## References and links

[1] A. D. Ellis, Z. Jian, and D. Cotter, "Approaching the Non-Linear Shannon Limit," Journal of Lightwave Technology **28**, 423-433 (2010).
[2] R. Essiambre, G. Kramer, P. J. Winzer, G. J. Foschini, and B. Goebel, "Capacity Limits of Optical Fiber Networks," Journal of Lightwave Technology **28**, 662-701 (2010).
[3] E. Ip and J. M. Kahn, "Compensation of Dispersion and Nonlinear Impairments Using Digital Backpropagation," Journal of Lightwave Technology **26**, 3416-3425 (2008).
[4] C. Xi, L. Xiang, S. Chandrasekhar, B. Zhu, and R. W. Tkach, "Experimental demonstration of fiber nonlinearity mitigation using digital phase conjugation," in *Optical Fiber Communication Conference and Exposition (OFC/NFOEC), 2012,* paper OTh3C.
[5] S. L. Jansen, D. Van den Borne, B. Spinnler, S. Calabro, H. Suche, P. M. Krummrich,W. Sohler, G. Khoe and H. de Waardt, "Optical phase conjugation for ultra long-haul phase-shift-keyed transmission," Journal of Lightwave Technology **24**, 54-64 (2006).
[6] D. M. Pepper and A. Yariv, "Compensation for phase distortions in nonlinear media by phase conjugation," Optics Letters **5**, 59-60 (1980).
[7] I. Phillips, M. Tan, M. F. Stephens, M. McCarthy, E. Giacoumidis, S. Sygletos*,* Pawel Rosa, Simon Fabbri, Son T. Le, Thavamaran Kanesan, Sergei K. Turitsyn, Nick J. Doran, Paul Harper and Andrew D. Ellis, "Exceeding the Nonlinear-Shannon Limit using Raman Laser Based Amplification and Optical Phase Conjugation," in *Optical Fiber Communication Conference*, San Francisco, California, 2014, paper M3C.1.
[8] S. Watanabe, S. Kaneko, and T. Chikama, "Long-Haul Fiber Transmission Using Optical Phase Conjugation," Optical Fiber Technology **2**, 169–178 (1996).
[9] A. R. C. X. Liu, P. J. Winzer, R. W. Tkach, and S. Chandrasekhar, "Phase-conjugated twin waves for communication beyond the Kerr nonlinearity limit," Nat. Photonics **7**, 560-568 (2013).

[10] G. P. Agrawal, *Fiber-Optic Communication Systems*, 4th ed. (Wyley, 2010).
[11] A. Hasegawa and Y. Kodama, *Solitons in Optical Communications* (Oxford University Press, 1996).
[12] V. E. Zakharov and A. B. Shabat, "Exact theory of two-dimensional self-focusing and one-dimensional self-modulation of waves in nonlinear media," Soviet Physics-JETP **34**, 62–69 (1972).
[13] L. F. Mollenauer and J. P. Gordon, *Solitons in Optical Fibers: Fundamentals and Applications* (Academic Press, 2006).
[14] Y. S. Kivshar and G. P. Agrawal, *Optical Solitons: From Fibers to Photonic Crystals* (Academic Press, 2003).
[15] A. Peleg, M. Chertkov, and I. Gabitov, "Inelastic interchannel collisions of pulses in optical fibers in the presence of third-order dispersion," J. Opt. Soc. Am. B **21**, 18-23 (2004).
[16] J. E. Prilepsky, S. A. Derevyanko, and S. K. Turitsyn, "Temporal Solitonic Crystals and Non-Hermitian Informational Lattices," Phys. Rev. Lett. **108**, 183902 (2012).
[17] O. V. Yushko and A. A. Redyuk, "Soliton communication lines based on spectrally efficient modulation formats," Quantum Electron **44**, 606-611 (2014).
[18] M. J. Ablowitz, D. J. Kaup, A. C. Newell, and H. Segur, "The inverse scattering transform-Fourier analysis for nonlinear problems," Stud. Appl. Math. **53**, 249–315 (1974).
[19] V. E. Zakharov, S. V. Manakov, S. P. Novikov, and L. P. Pitaevskii, *Theory of Solitons. The Inverse Scattering Method.* (Colsultants Bureau, New York, 1984).
[20] M. J. Ablowitz and H. Segur, *Solitons and the Inverse Scattering Transform.* (SIAM, Philadelphia, 1981).
[21] A.R. Osborne, "The inverse scattering transform: tools for the nonlinear Fourier analysis and filtering of ocean surface waves," Chaos, Solitons & Fractals **5**, 2623-2637 (1995).
[22] A.S. Fokas and I. M. Gelfand. "Integrability of linear and nonlinear evolution equations and the associated nonlinear Fourier transforms," Letters in Mathematical Physics **32**, 189-210 (1994).
[23] A. Hasegawa and T. Nyu, "Eigenvalue communication," Journal of Lightwave Technology **11**, 395-399 (1993).
[24] E. G. Turitsyna and S. K. Turitsyn, "Digital signal processing based on inverse scattering transform," Opt. Lett. **38**, 4186-4188 (2013).
[25] H. Terauchi and A. Maruta, "Eigenvalue Modulated Optical Transmission System Based on Digital Coherent Technology," in *18th OptoElectronics and Communications Conference held jointly with 2013 International Conference on Photonics in Switching (OECC/PS), IEICE, 2013,* paper WR2-5.
[26] M. I. Yousefi and F. R. Kschischang, "Information transmission using the nonlinear Fourier transform, Part I: Mathematical tools," IEEE Trans. Inf. Theory **60**, 4312 - 4328 (2014).
[27] M. I. Yousefi and F. R. Kschischang, "Information transmission using the nonlinear Fourier transform, Part II: Numerical methods," IEEE Trans. Inf. Theory **60**, 4329-4345 (2014).
[28] M. I. Yousefi and F. R. Kschischang, "Information transmission using the nonlinear Fourier transform, Part III: Spectrum modulation," IEEE Trans. Inf. Theory **60**, 4346- 4369 (2014).
[29] J. E. Prilepsky, S. A. Derevyanko, and S. K. Turitsyn, "Nonlinear spectral management: Linearization of the lossless fiber channel," Optics Express **21**, 24344-24367 (2013).
[30] J. E. Prilepsky, S. A. Derevyanko, K. J. Blow, I. Gabitov, and S. K. Turitsyn, "Nonlinear inverse synthesis and eigenvalue division multiplexing in optical fiber channels," Phys. Rev. Lett. **113**, 013901 (2014).
[31] S. Hari, F. Kschischang, and M. Yousefi, "Multi-eigenvalue communication via the nonlinear Fourier transform," in *Communications (QBSC), 2014, 27th Biennial Symposium on*, pp. 92-95.
[32] S. Chandrasekhar and L. Xiang, "OFDM Based Superchannel Transmission Technology," *J*ournal of Lightwave Technology **30**, 3816-3823 (2012).
[33] E. Giacoumidis, M. A. Jarajreh, S. Sygletos, S. T. Le, F. Farjady, A. Tsokanos*,* A. Hamié, E. Pincemin, Y. Jaouën, A. D. Ellis and N. J. Doran, "Dual-polarization multi-band optical OFDM transmission and transceiver limitations for up to 500 Gb/s uncompensated long-haul links," Optics Express **22**, 10975-10986 (2014).
[34] J. Armstrong, "OFDM for Optical Communications," Journal of Lightwave Technology **27**, 189-204 (2009).
[35] S.B. Weinstein, "The history of orthogonal frequency-division multiplexing [history of communications]," IEEE Commun. Mag **47,** 26-35 (2009).
[36] W. Shieh, C. Athaudage, "Coherent optical orthogonal frequency division multiplexing," Electron. Lett. **42**, 587-588 (2006).
[37] S. T. Le, K. Blow, and S. Turitsyn, "Power pre-emphasis for suppression of FWM in coherent optical OFDM transmission," Optics Express **22**, 7238-7248 (2014).
[38] D. Hillerkuss, R. Schmogrow, M. Meyer, S. Wolf, M. Jordan, P. Kleinow, N. Lindenmann, P. Schindler, A. Melikyan, X. Yang, S. Ben-Ezra, B. Nebendahl, M. Dreschmann, J. Meyer, F. Parmigiani, P. Petropoulos, B. Resan, A. Oehler, K. Weingarten, L. Altenhain, T. Ellermeyer, M. Moeller, M. Huebner, J. Becker, C. Koos, W. Freude, and J. Leuthold, "Single-Laser 32.5Tbit/s Nyquist WDM Transmission," Journal of Optical Communications and Networking **4**, 715-723 (2012).
[39] A. C. Newell, *Solitons in mathematics and physics* (SIAM Philadelphia, 1985).
[40] R. Bulirsch, *Introduction to Numerical Analysis, 2nd ed.* (Springer-Verlag, 1993).
[41] S. Burtsev, R. Camassa, and I. Timofeyev, "Numerical Algorithms for the Direct Spectral Transform with Applications to Nonlinear Schrödinger Type Systems," Journal of Computational Physics **147**, 166-186 (1998).
[42] G. Boffetta and A. R. Osborne, "Computation of the direct scattering transform for the nonlinear Schroedinger equation," Journal of Computational Physics **102**, 252-264 (1992).
[43] O. V. Belai, L. L. Frumin, E. V. Podivilov, and D. A. Shapiro, "Efficient numerical method of the fiber Bragg grating synthesis," Journal of the Optical Society of America B **24**, 1451-1457 (2007).

[44] T Kuusela, J Hietarinta, K Kokko, and R Laiho, "Soliton experiments in a nonlinear electrical transmission line," European Journal of Physics **8**, 27-33 (1987).
[45] S. Zohar, "Toeplitz Matrix Inversion: The Algorithm of W. F. Trench," Journal of the Association for Computing Machinery **16**, 592-561 (1969).
[46] M. V. Barel, G. Heinig, and P. Kravanja, "A Stabilized Superfast Solver for Nonsymmetric Toeplitz Systems," SIAM J. Matrix Anal. Appl. **23**, 494–510 (2001).
[47] M. Stewart, "Fast algorithms for structured matrix computations" in *Handbook of Linear Algebra* (2$^{nd}$ edition) (Chapman & Hall, 2013), Chap. 62.
[48] A. Buryak, J. Bland-Hawthorn, and V Steblina, "Comparison of Inverse Scattering Algorithms forDesigning Ultrabroadband Fibre Bragg Gratings," Opt. Express **17**, 1995–2004 (2009).
[49] G.Xiao and K. Yashiro, "An Efficient Algorithm for Solving Zakharov–Shabat Inverse Scattering Problem," IEEE Trans. Antennas and Propagation **50**, 807-811 (2002).
[50] S. Wahls and H. V. Poor, "Introducing the fast nonlinear Fourier transform," in *Proceedings of International Conference on Acoustics, Speech, and Signal Processing (ICASSP 2013), IEEE*, 2013, pp. 5780–5784.
[51] S. Wahls and H. V. Poor, "Fast Numerical Nonlinear Fourier Transforms," submitted to IEEE Trans. Inf. Theory (2013)*,* http://arxiv.org/abs/1402.1605.
[52] S.V. Manakov, "On the theory of two-dimensional stationary self-focusing of electromagnetic waves", Soviet Physics-JETP **38**, 248-253 (1974).
[53] R. A. Shafik, S. Rahman, and R. Islam, "On the Extended Relationships Among EVM, BER and SNR as Performance Metrics," in *Electrical and Computer Engineering, 2006. ICECE '06. International Conference on*, 2006, pp. 408-411.
[54] S. T. Le, K. J. Blow, V. K. Menzentsev, and S. K. Turitsyn, "Comparison of numerical bit error rate estimation methods in 112Gbs QPSK CO-OFDM transmission," in *Optical Communication (ECOC 2013), 39th European Conference and Exhibition on*, 2013, paper P4.14.
[55] S. Kilmurray, T. Fehenberger, P. Bayvel, and R. I. Killey, "Comparison of the nonlinear transmission performance of quasi-Nyquist WDM and reduced guard interval OFDM," Optics Express **20**, 4198-4205 (2012).
[56] Y. Lu, Y. Fang, B. Wu, K. Wang, W. Wan, F. Yu, L. Li, X. Shi and Q. Xiong, "Experimental comparison of 32-Gbaud Electrical-OFDM and Nyquist-WDM transmission with 64GSa/s DAC," in *Optical Communication (ECOC 2013), 39th European Conference and Exhibition on* (2013).

## 1. Introduction

The rapidly increasing demand on communication speed is exerting great pressure on the networks' infrastructure at every scale, which explains the real motivation behind all optical communications research. Since the introduction of fiber-optic communications in the late 1970s, many technological advances, such as erbium-doped fiber amplifiers, wavelength division multiplexing (WDM), dispersion management, forward error correction, and Raman amplification, have been developed to enable the exponential growth of data traffic [1]. However, the continuing bandwidth demand is pushing the required capacity close to the theoretical limit of the standard single-mode fiber (SSFM), which is imposed by the fiber's nonlinearity effects (Kerr effect) [2]. In recent years, there have been extensive efforts in attempting to surpass the Kerr nonlinearity limit through various nonlinearity compensation techniques, including digital back-propagation (DBP) [3], digital [4] and optical [5-7] phase conjugations (OPCs) at the mid-link or installed at the transmitter [8], and phase-conjugated twin waves [9]. However, there are still many limitations and challenges in applying the aforementioned nonlinear compensation methods, because the transmission technologies utilized in optical fiber communication systems were originally developed for linear (radio or open space) communication channels. Therefore, it would constitute a significant development if the fiber nonlinearity could be taken into account in a "constructive way" when designing the core optical communication coding, transmission, detection, and processing approaches. In simple terms, the true limits of nonlinear fiber channels are yet to be found.

The propagation of optical signals in fiber can be accurately modelled by the nonlinear Schrödinger equation. (NLSE) [10], which describes the continuous interplay between dispersion and nonlinearity. It is well known that the NLSE (without perturbation) belongs to the class of integrable nonlinear systems [11, 12]. In particular, this means that the NLSE allows the existence of a special type of solutions: highly robust nonlinear waves, called solitons. Solitons were proposed as the information carriers for the high-capacity fiber-optic communications [11, 13, 14]. However, on-off keying transmission built on pure fundamental solitons (i.e., where the solitons have been used as individual information-bearing pulses) was affected by the soliton collision problems and ensuing limitations due to the inter-channel cross-talk in wavelength division multiplexing (WDM) lightwave transmission systems [11, 15], leading to the reduction in spectral efficiency; although some recent works demonstrated positive results with regard to coherent soliton-based transmission using multilevel modulation [16, 17]. Another, less known in the optical communication community, consequence of the NLSE's integrability is that it allows one to present the field evolution within a special basis of *nonlinear normal modes*, including non-dispersive soliton modes and quasilinear dispersive radiation. The dynamics of individual nonlinear modes is essentially linear, which means that the nonlinearity-induced cross-talk between these modes is effectively absent during the propagation [18-22]. From an information and communication theory point of view, these nonlinear modes can potentially be used to encode information that, in turn, can be recovered at the receiver without suffering from the nonlinear impairments [12, 18-22]. The prefiguration of this general idea was first introduced by Hasegawa and Nyu in [23], termed as *eigenvalue communications*. In the original version [23], this approach was focused on the invariance of the discrete eigenvalues (i.e., those attributed to the solitonic degrees of freedom) of the Lax operator associated with the NLSE, that were further used to encode and transmit information. In more general words, the class of digital processing approaches based on the integrability of the underlying dynamical system (at least in the leading approximation) opens fundamentally new possibilities for advanced modulation, coding, and transmission schemes, which are inherently resistant against the nonlinear fiber effects. The implementation of this research program leads to the foundation of a nonlinear communication theory.

The solution of an integrable equation can be determined by applying the inverse scattering transform (IST) technique, which serves as a nonlinear analogue of the Fourier transform (FT) approach applied for the solution of a linear partial differential equation [12,18-22]; thus, the Nonlinear Fourier transform (NFT) term is quite frequently used for IST operations [18-22]. The IST method involves three stages, each having its counterpart in the linear Fourier method. i) First, one decomposes the given input waveform into scattering data, which requires the solution of the so-called spectral Zakharov–Shabat problem (ZSP); this operation serves the same role as the forward FT does for a linear system, and therefore one can refer to it as to the forward NFT (FNFT). The FNFT represents the signal in terms of its continuous and discrete spectral data that correspond to the non-soliton (radiative dispersive) and soliton (non-dispersing) parts of the NLSE solution, respectively. ii) The second stage of the IST method involves the propagation of the spectral data to a desired distance, and this propagation *is trivial and linear,* which constitutes the core of the IST-based transmission concept: The solitonic eigenvalues are integrals of motion and do not change at all, and the continuous nonlinear spectrum obeys the dispersion law of the linearized NLSE [12,18-23]. iii) The final stage, which accomplishes the solution of the NLSE at a given distance, involves the solution of the Gelfand–Levitan–Marchenko equation (GLME), providing the recovery of the field profile in the space-time domain from the scattering data [12,18-23]; we will refer to this operation as to the backward NFT (BNFT).

Insofar as the evolution of both continuous and discrete spectra in a lossless fiber channel is linear, they can both be used for encoding and transmitting information, being formally completely insensitive to fiber nonlinearity. This feature opens the way for effectively linear signal propagation trough a nonlinear fiber. As a result, there has recently been a considerable renewed interest in transmission based on the modulation of the nonlinear spectral data and in the utilization of IST-based processing [24-31]. In particular, in [23,26-28,31] the discrete (solitonic) components of the nonlinear Fourier spectrum were studied for data communications. In [31] it was shown that the multi-eigenvalue pulses can be designed to achieve a spectral efficiency (SE) greater than 1 bits/s/Hz. However, this approach requires considerable optimization of the pulse shapes for the purpose of maximizing the resulting SE.

In addition, multiple pulse shapes are required to encode the transmitted information, so that more than 2 million different pulse shapes were studied in [31] to allow a potential SE of 3.14 bits/s/Hz to be achieved. In general, the transmission approach based on the encoding of information on the discrete components of the nonlinear Fourier spectrum is still at an early stage and needs further efforts for efficient implementation. Moreover, in [26-28,31] only the propagation of a single pulse was considered to demonstrate the performance of the approach and a complete transmission system has yet to be explored.

An alternative and effective approach for achieving a high SE in communication systems based on the NFT has been proposed in [24,29,30] by exploiting the vast amount of available degrees of freedom contained in the continuous part of the nonlinear spectrum; in [24] this method was applied to the NLSE with normal chromatic dispersion, which is somewhat simpler than the anomalous dispersion case considered in our present study, as the solitons there do not emerge from a finite input. In this study, we utilize the transmission based on the nonlinear inverse synthesis (NIS) [30], a method which allows one to avoid the problems associated with the solitary degrees of freedom in the anomalous dispersion NLSE. Within this approach, the encoded input signal (or, to be more specific, its linear spectrum) is first mapped onto a complex field in the time domain via the GLME (i.e., via the BNFT) before transmission (see the detailed flowchart presented in Fig. 1 of [30]). Importantly, the input field profile obtained via the GLME is soliton-free, meaning that there is no soliton part contained in the nonlinear spectrum as the total number of solitons is the conserved quantity for the NLSE evolution. At the receiver, the nonlinear spectrum of the complex field can be obtained by solving the ZSP (i.e., by the FNFT) and then the transmitted continuous spectrum can be recovered using a single-tap linear dispersion removal; the latter step is needed solely for the recovery of the initial phase information and it is possible to use a modulation format where this step is unnecessary.

Since the input field for the BNFT operation (GLME) can be arbitrary, different modulation formats can be combined with the NIS method, providing the flexibility in the system's design for achieving a high SE and potentially even mitigation of the corruptions occurring because of the deviation of the true signal evolution from the pure NLSE model. In addition, the proposed NIS approach, which is a DSP-based approach, can also be easily integrated with the current coherent transmission technology. Finally, the numerical complexity of NIS can be competitive and potentially even outperform that of the digital back propagation (DBP) based methods [24,29,30] (see also Subsection 4.3 below).

In this paper, we essentially extend the preliminary idea from [30] by designing and investigating the performance of high-SE communication systems based on the NIS approach. We consider orthogonal frequency division multiplexing (OFDM) and Nyquist-shaped 112 Gb/s, 224 Gb/s, and 336 Gb/s coherent transmission systems with QPSK, 16QAM, and 64QAM modulation formats. To show the potential benefit of this approach, we also compare the performance of the NIS system with the multi-step per span DBP. Our Monte Carlo simulations show that with a proper design, the high-SE NIS-based communication system can offer a performance improvement of up to 4.5 dB, which is comparable with that achievable with the multi-step per span DBP. Additionally, we discuss the challenges and current limitations of the NIS approach for high-capacity optical communications.

## 2. Nonlinear inverse synthesis for transmission in optical fiber

In this section, we present the basic block functions of an NIS-based coherent optical communication system, which are shown in Fig. 1(a). Details of these block functions will be discussed in Section 3.

At the transmitter, the transmitted binary data sequence is first digitally encoded onto a complex waveform ($s(t)$) using an arbitrary modulation format and coding technique. The linear Fourier spectrum of this encoded complex waveform is defined as:

$$S(\omega) = \int_{-\infty}^{\infty} s(t)\exp(-j\omega t)dt \qquad (1)$$

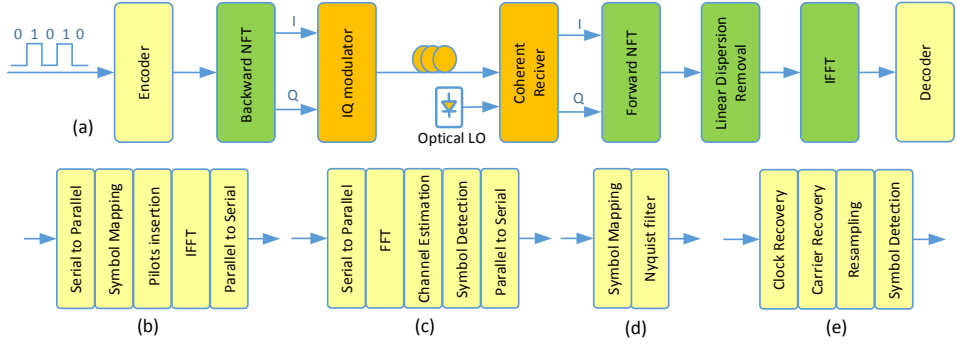

Fig. 1 (a) Block diagram of NIS-based optical communication systems, (b, c) block diagrams of OFDM encoder and decoder, (d, e) block diagrams of Nyquist-shaped encoder and decoder.

After the encoder, the linear Fourier spectrum of the encoded complex waveform, $S(\omega)$, is mapped onto the continuous part of the nonlinear spectrum of a complex signal $q(t)$ using the BNFT (i.e., by solving the GLME). Let $r(\xi)$ denote the continuous part of the nonlinear spectrum of $q(t)$ (see the definitions in Sections 3.2, 3.3). Then, the mapping operation of the BNFT block can be expressed as (c.f. Eq. (16) below):

$$r(\xi)\big|_{\xi=-\omega/2} = -S(\omega) \qquad (2)$$

The complex signal, $q(t)$, is then fed into the IQ modulator for direct up-converting into the optical domain and launched into the fiber. We note that the only additional DSP block required in the transmitter of a NIS-based system is the BNFT block, which maps the encoded information onto the continuous part of the nonlinear spectrum of a complex signal. The complex transmitted signal (output of the BNFT block) in this case is soliton-free, meaning that the nonlinear spectrum contains no discrete part. At the receiver, the real and imaginary parts of the transmitted signal are detected with a coherent receiver. The nonlinear spectrum of the received signal is obtained by using the forward NFT (i.e. by solving the ZSP). After the propagation over a lossless optical fiber channel (e.g., with the distributed Raman amplification), the evolution of $r(\xi)$ is trivial [11,12]:

$$r(L,\xi) = r(\xi) \cdot e^{-2j\xi^2 L} \qquad (3)$$

where $L$ is the transmission distance. Expression (3) basically shows that the orthogonality of nonlinear normal modes is preserved during the signal propagation within the NLSE. This property allows us to recover the linear Fourier spectrum of the initial encoded complex signal after the decoder by applying a single step linear dispersion removal:

$$\overline{S}(\omega) = -r(L,\xi) \cdot e^{2j\xi^2 L} \big|_{\xi=-\omega/2} \qquad (4)$$

Then, having performed the dispersion removal, the complex waveform $s(t)$ can be recovered using the IFFT operation and, finally, it can be fed into the standard decoder for data detection.

As described above, the DSP at the receiver of an NIS-based system involves solving the ZSP and a single linear compensation step to remove the nonlinear impairments without reverse propagation. This clearly demonstrates the advantage of the NIS method over the other nonlinear compensation techniques.

As mentioned above, the NIS method can be combined with any modulation formats and transmission schemes. In this work, we consider two transmission schemes, which have attracted a great amount of attention recently because of the possibility to achieve the Nyquist limit (1 symbol/s/Hz): the OFDM and the Nyquist-shaped modulations [32-38]. The simplified encoders and decoders for the OFDM and Nyquist-shaped systems are shown in Fig. 1(b), Fig. 1(c), Fig. 1(d), and Fig. 1(e), respectively. In the OFDM system, the information to be transmitted is encoded in the frequency domain onto a large number of overlapping orthogonal subcarriers. The orthogonality among the subcarriers allows the encoded information to be detected without suffering from intercarrier interference, even though the subcarriers overlap [34]. In optical communications, OFDM provides several advantages against other modulation formats, such as its improved tolerance to chromatic dispersion and a simplified equalization scheme [34]. For Nyquist-shaped systems, the information is encoded directly in the time domain using a Nyquist filter (analog or digital) to generate a sinc-shaped pulse. The sinc pulse not only satisfies the intersymbol-free condition but also provides a rectangular spectrum, allowing one to achieve the Nyquist limit [38]. As a result, the performance comparison of the OFDM and Nyquist-shaped modulations of the nonlinear spectrum combined with the NIS method is of great specific interest with regard to the high-SE transmission.

## 3. Channel model and NFT basics

### 3.1. NLSE and normalizations

We consider as a master model the NLSE governing the propagation of a complex slow-varying optical-field envelope $q(z,t)$ along a single-mode lossless optical fiber [11-13]:

$$jq_z - \frac{\beta_2}{2} q_{tt} + \gamma q |q|^2 = 0 \qquad (5)$$

where $z$ stands for the propagation distance and $t$ is the time in the frame co-moving with the group velocity of the envelope. Here, we focus on the case of anomalous dispersion (i.e., the constant chromatic dispersion coefficient is $\beta_2 < 0$ in Eq. (5)) and hence deal with the so-called focusing type of NLSE. The higher-order dispersion terms are not considered here. The instantaneous Kerr nonlinearity coefficient $\gamma$ is expressed through the nonlinear part of the refractive index $n_2$ and an effective mode area $A_{eff}$:

$$\gamma = n_2 \omega_0 / c A_{eff} \qquad (6)$$

where $c$ is the vacuum speed of light and $\omega_0$ is the angular carrier frequency.
When dealing with the NFT, it is convenient to work with the NLSE in the normalized form:

$$jq_z + \frac{1}{2} q_{tt} + q|q|^2 = 0, \qquad (7)$$

which can be obtained through the following rescaling of variables:

$$\frac{t}{T_s} \to t, \quad \frac{z}{Z_s} \to z, \quad q\sqrt{\gamma Z_s} \to q \qquad (8)$$

where $T_s$ is a free parameter (e.g., a characteristic time scale of the input waveform) and the associated space scale is $Z_s = T_s^2 / \beta_2$. Note that all the quantities in the normalized Eq. (7), namely, $q$, $t$, and $z$, are now dimensionless.

### 3.2. Forward NFT as part of the IST method.

Similarly to the forward FT, the purpose of the FNFT is to decompose the signal into the IST spectral data. In other words, the forward NFT converts a complex signal into its correspondent nonlinear spectrum, including both continuous and discrete parts. This operation is achieved by solving the ZSP [12]. The latter corresponds to a scattering problem for a non-Hermitian (in the case of anomalous dispersion) Dirac-type system of equations for two auxiliary functions $v_{1,2}(t)$, with the NLSE input pulse profile, $q(0,t) \equiv q(t)$, serving as an effective potential entering into these equations:

$$\frac{dv_1}{dt} = q(t)v_2 - j\varsigma v_1, \quad \frac{dv_2}{dt} = -\bar{q}(t)v_1 + j\varsigma v_2 \qquad (9)$$

Here, $\zeta$ is a (generally complex) eigenvalue, $\zeta = \xi + j\eta$, $\bar{q}(t)$ is the complex conjugation of the potential $q(t)$, which is assumed to decay as $t \to \pm\infty$ (for the exact conditions imposed on the decay rate, see [11,12]).

In order to define the continuous part of the nonlinear spectrum (for real $\zeta = \xi$) one fixes two linearly-independent Jost solutions of Eq. (9) as:

$$\Phi(t,\xi) = [\phi_1, \phi_2]^T, \quad \widetilde{\Phi}(t,\xi) = [\bar{\phi}_2, -\bar{\phi}_1]^T, \qquad (10)$$

with the "initial" condition at the left end:

$$\Phi\big|_{t\to-\infty} = \left[e^{-j\xi t}, 0\right]^T \qquad (11)$$

In the same way, we fix two other Jost solutions,

$$\Psi(t,\xi) = [\psi_1, \psi_2]^T, \quad \widetilde{\Psi}(t,\xi) = [\bar{\psi}_2, -\bar{\psi}_1]^T, \qquad (12)$$

at the right end:

$$\Psi\big|_{t\to+\infty} = \left[0, e^{j\xi t}\right]^T \qquad (13)$$

These two solution sets are linearly dependent and can be expressed through the Jost scattering coefficients $a(\xi)$ and $b(\xi)$ as:

$$\begin{cases} \Phi(t,\xi) = a(\xi)\widetilde{\Psi}(t,\xi) + b(\xi)\Psi(t,\xi) \\ \widetilde{\Phi}(t,\xi) = -\bar{a}(\xi)\Psi(t,\xi) + \bar{b}(\xi)\widetilde{\Psi}(t,\xi) \end{cases} \qquad (14)$$

The (left) reflection coefficient (this term corresponds to the notions adopted within the inverse scattering transform theory and does not refer to either "left" or "right" end of the fiber, so that the word "reflection" must not be confused with any material property), giving the continuous part of the nonlinear spectrum, is defined as:

$$r(\xi) = \bar{b}(\xi)/a(\xi), \qquad (15)$$

and the solitons correspond to the complex eigenvalues $\zeta_n$, where $a(\zeta_n) = 0$. In general, the forward NFT maps the initial field, $q(0,t)$, onto a set of scattering data $\Sigma = [(r(\xi), \xi$ is real); $(\zeta_n, \gamma_n = (b(\zeta_n)a'_\zeta(\zeta_n))^{-1})]$, where the index $n$ runs over all discrete eigenvalues of the ZSP (discrete non-dispersive part of the nonlinear spectrum). Herein, the complex-valued function $r(\xi)$ of the real argument $\xi$ (nonlinear spectrum) is similar to the usual Fourier spectrum. Therefore, within the NIS method we use this continuous part of the nonlinear spectrum, $r(\xi)$, to encode and transmit the information, and $\xi$ plays the role of the frequency. The nonlinear spectral (NS) function, defined as:

$$N(\omega) = -r(\xi)\big|_{\xi=-\omega/2} \qquad (16)$$

serves as the direct nonlinear analog of the Fourier spectrum, tending to the ordinary FT of $q(0,t)$ in the linear limit (see [29] for explicit examples).

*3.3. BNFT as a part of the IST method*

The BNFT is an inverse operation for the forward NFT that maps the scattering data $\Sigma$ onto the field $q(t)$ (or, more generally, onto $q(t,z)$) in the time domain. This can be achieved via the GLME for the unknown function $K(t, x)$ [18-20,39]. In the NIS method described above, we consider only the soliton-free case, for which the GLME can be written as:

$$K(t,x) + F(t+x) + \int_{-\infty}^{t}\int_{-\infty}^{t} K(t,\lambda)\overline{F}(\lambda+\sigma)F(\sigma+x)d\sigma d\lambda, \qquad (17)$$

Here, $F(t)$ is the linear FT of $r(\xi)$, given by

$$F(t) = \frac{1}{2\pi}\int_{-\infty}^{+\infty} r(\xi)e^{-j\xi t}d\xi \qquad (18)$$

After solving Eq. (17) for $K(t, x)$, the inverse NFT of $r(\xi)$ can eventually be obtained as:

$$q(t) = 2\lim_{x \to t-0} K(t,x) \qquad (19)$$

In the following section, we will discuss numerical methods for computing the forward and backward NFTs.

## 4. Numerical methods

*4.1. Numerical methods for computing the continuous spectrum*

The continuous spectrum (i.e., Jost coefficients $a(\xi)$ and $b(\xi)$ and the corresponding $r(\xi)$, see Eq.(15)) can be computed by directly integrating the Zakharov–Shabat system (9) and then evaluating the limits for the corresponding Jost function components as:

$$a(\xi) = \lim_{t \to +\infty} \phi_1(t,\xi)e^{j\xi t}, \quad b(\xi) = \lim_{t \to +\infty} \phi_2(t,\xi)e^{-j\xi t} \qquad (20)$$

Several discretization and integration methods have been proposed to solve the ZSP (9), including the forward and center discretizations with first-order Euler method [40], the fourth-order Runge–Kutta method [41], and the piecewise-constant approximation (PCA) method [42,43]. From an implementation point of view, the PCA method offers an attractive solution because it can be implemented effectively in parallel to reduce the computational time [42,43]. In addition, despite the fact that the Runge–Kutta method is of a higher order, the extra accuracy can be lost because of the additional dependence of its numerical error on the eigenvalue [42]. This phenomenon limits the usefulness of the Runge–Kutta algorithm to a region inside the unit circle around the origin in the complex plane of the eigenvalues. Therefore, only the PCA method will be considered in our paper. The other numerical methods for computing the FNFT are discussed in [27] and in the references therein.

Although the ZSP is defined on the infinite time line, we must truncate the potential outside a sufficiently large interval in order to make the numerical solution possible. As a result, we reduce the infinite-line spectral problem to a problem with a finite -width potential and to the corresponding boundary conditions for the truncated potential.

Now, we recall the basic elements of the PCA, which is conceptually a variant of a layer peeling algorithm applied for the solution of the ZSP [42]. The potential $q(t)$ is truncated outside a sufficiently large interval $(-T_0; T_0)$. Inside this interval, $q(t)$ is chosen to be constant, $q_n=q(t_n)$, on each elementary subinterval (or numerical time-step) $(t_n-\Delta t/2; t_n+\Delta t/2)$, where $t_n = -T_0+n\Delta t$, $\Delta t=T_0/M$ is the time step, and $2M+1$ is the total number of discretization points inside the considered truncation interval. The idea of the PCA method is based on the fact that Eq. (9) can be solved exactly inside each elementary subinterval for an arbitrary value of the spectral parameter $\xi$ as:

$$\Phi(t_n + \Delta t/2, \xi) = T(q_n, \xi)\Phi(t_n - \Delta t/2, \xi), \tag{21}$$

where the transfer matrix $T(q_n, \xi)$ is given by

$$\begin{aligned}T(q_n, \xi) &= \exp\left[\Delta t\begin{pmatrix}-j\xi & q_n \\ -q_n^* & j\xi\end{pmatrix}\right] \\ &= \begin{pmatrix}\cosh(k\Delta t) - j\xi k^{-1}\sinh(k\Delta t) & q_n k^{-1}\sinh(k\Delta t) \\ -q_n^* k^{-1}\sinh(k\Delta t) & \cosh(k\Delta t) + j\xi k^{-1}\sinh(k\Delta t)\end{pmatrix},\end{aligned} \tag{22}$$

Here, $k = j\sqrt{q_n^2 + \xi^2}$ is a constant parameter in each interval $\Delta t$.

The scattering problem can be solved by "propagating" the solution iteratively, starting from $-T_0$ towards the right truncation border $T_0$, using the set of transfer matrices $T(q_n, \xi)$ given by Eq. (22). The final result can be expressed as:

$$\Phi(T_0 - \Delta t/2, \xi) = \Pi(\xi)\Phi(-T_0 - \Delta t/2, \xi),$$
$$\Pi(\xi) = \prod_{n=1}^{2M} T(q_n, \xi) \tag{23}$$

The initial condition (13) defined at the right truncation end can be written as:

$$\Phi(-T_0 - \Delta t/2, \xi) = \begin{pmatrix}1 \\ 0\end{pmatrix}e^{-j\xi(-T_0 - \Delta t/2)} \tag{24}$$

Then, at the left end of the full interval we have:

$$\begin{aligned}\Phi(T_0 - \Delta t/2, \xi) &= \begin{pmatrix}a(\xi)e^{-j\xi(T_0 - \Delta t/2)} \\ b(\xi)e^{j\xi(T_0 - \Delta t/2)}\end{pmatrix} = \begin{pmatrix}\Pi_{11}(\xi) & \Pi_{12}(\xi) \\ \Pi_{21}(\xi) & \Pi_{22}(\xi)\end{pmatrix}\Phi(-T_0 - \Delta t/2, \xi) \\ &= \begin{pmatrix}\Pi_{11}(\xi) & \Pi_{12}(\xi) \\ \Pi_{21}(\xi) & \Pi_{22}(\xi)\end{pmatrix}\begin{pmatrix}e^{-j\xi(-T_0 - \Delta t/2)} \\ 0\end{pmatrix},\end{aligned} \tag{25}$$

and, therefore, the Jost coefficients are given by:

$$a(\xi) = \Pi_{11}(\xi)e^{2j\xi T_0}, \quad b(\xi) = \Pi_{21}(\xi)e^{-j\xi\Delta t} \tag{26}$$

In general, when the potential $q(t)$ is truncated outside the interval $(T_{min}, T_{max})$ with arbitrary borders, the expression (26) can be modified as:

$$a(\xi) = \Pi_{11}(\xi)e^{j\xi(T_{max} - T_{min})}, \quad b(\xi) = \Pi_{21}(\xi)e^{j\xi(T_{max} + T_{min} - \Delta t)} \tag{27}$$

Once can notice that the PCA method has some interesting similarities to the transmission line theory, in which the potential $q(t)$ can be considered as the distributed parameter of the line [44]. From (22), one can see that the transfer matrixes $T(q_n, \xi)$ can be calculated independently of each other. As a result, the PCA algorithm can be easily implemented in parallel to reduce the computational time for high-speed NIS-based systems.

To demonstrate the performance of the PCA method, we consider here a rectangular pulse as an example (see, e.g., [27,29]):

$$q(t) = \begin{cases}A, & t \in [T_1, T_2] \\ 0 & \text{otherwise}\end{cases} \tag{28}$$

The continuous spectrum of this rectangular pulse is given by:

$$r(\xi) = \frac{\overline{A}}{j\xi} e^{-2j\xi t} \left( 1 - \frac{\sqrt{\xi^2 + |A|^2}}{j\xi} \cot\left(\sqrt{\xi^2 + |A|^2}\,(T_2 - T_1)\right) \right) \qquad (29)$$

In Figs. 2(a) and 2(b) we compare the continuous spectra of the rectangular pulse with different amplitudes, which are calculated using the analytical formula (29) and the numerical PCA method. We define the normalized root mean squared error (RMSE) of the PCA as follows:

$$MSE = \frac{1}{A}\sqrt{\frac{1}{N}\sum_{k=1}^{N}|r_{exact}(\xi_k) - r_{numeric}(\xi_k)|^2} \qquad (30)$$

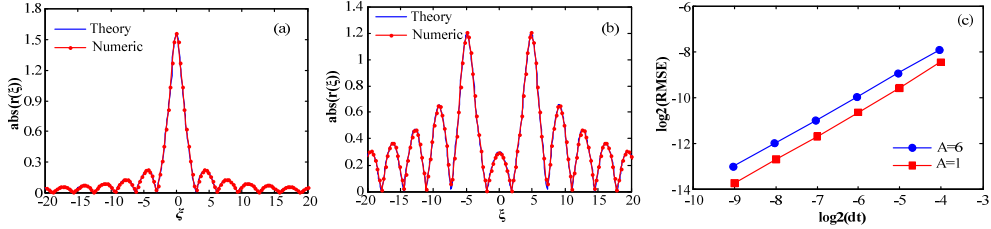

Fig. 2 (a) Continuous spectrum of rectangular pulse for $A = 1$, $T_2 - T_1 = 1$. (b) Continuous spectrum of rectangular pulse for $A = 6$, $T_2 - T_1 = 1$. (c) Mean square error as a function of the simulation time resolution ($dt$).

The dependence of the normalized RMSE on the time resolution $dt$ is shown in Fig. 2(c). One can note that the RMSE of the PCA method depends on the pulse amplitude or, in general, on the total signal energy. This is a fundamental limitation of PCA and other numerical methods for calculating the forward NFT. We will discuss this issue in more detail in the following sections.

*4.2. Numerical method for computing the BNFT*

The GLME (17) can be rewritten in the form of two coupled integrated Eqs. [49], which are more convenient for numerical calculations:

$$A_1(x,t) + \int_{-\infty}^{x} F(t+y)\overline{A}_2(x,y)dy$$
$$A_2(x,t) - \int_{-\infty}^{x} F(t+y)\overline{A}_1(x,y)dy = F(x+t), \quad x > t \qquad (31)$$

where $F(t)$ is the linear backward FT of $r(\xi)$ as given by Eq. (18).
For the numerical analysis we use the following change of the variables [43]:

$$u(x,s) = \overline{A}_1(x, x-s), \quad v(x,\tau) = \overline{A}_2(x, \tau - x), \qquad (32)$$

and then the GLME in the form (31) can be rewritten as:

$$u(x,s) + \int_{s}^{2x} \overline{F}(\tau - s)v(x,\tau)d\tau = 0,$$
$$v(x,\tau) - \int_{0}^{\tau} F(\tau - s)u(x,s)ds = F(\tau), \qquad (33)$$

Functions $u(x,s)$ and $v(x,s)$ are defined inside the interval $0 \leq \tau \leq 2x \leq 2T_0$. The BNFT mapping $r(\xi)$ onto the time domain is then given by

$$q(x) = 2v(x, 2x - 0) \tag{34}$$

Following the discretization procedure provided in [43], we divide the interval $0 \leq \tau \leq 2T_0$, where the function $F(\tau)$ is known, into segments of length $h=2T_0/N$. The discrete variables $\tau_n$, $s_k$, and $x_m$ are defined as:

$$\begin{aligned} s_k &= h(k - 1/2), & k &= 1, 2...m \\ \tau_n &= h(n - 1/2), & n &= 1, 2...m \\ x_m &= mh/2, & m &= 1, 2...N \end{aligned} \tag{35}$$

We also define the grid functions:

$$u_n^{(m)} = u(x_m, \tau_n), \quad v_n^{(m)} = v(x_m, \tau_n), \quad F_n = F(nh) \tag{36}$$

Using the rectangular quadrature scheme to approximate the integrals in Eq. (33), one obtains the following discrete form of the GLME:

$$\begin{aligned} u_k^{(m)} + h \sum_{n=k}^{m} \overline{F}_{n-k} v_n^{(m)} &= 0, \\ v_n^{(m)} - h \sum_{k=1}^{m} F_{n-k} u_k^{(m)} &= F_n \end{aligned} \tag{37}$$

The $m^{th}$ mesh element of the BNFT (in the time domain) is then given by:

$$q^{(m)} = 2v_m^{(m)} \tag{38}$$

Equations (38) can now be written in a matrix form as:

$$\mathbf{G}^{(m)} \begin{pmatrix} u^{(m)} \\ v^{(m)} \end{pmatrix} = \mathbf{b}^{(m)}, \tag{39}$$

where $\mathbf{b}^{(m)}$ is formed from the zero vector of dimension $m$ and the vector of dimension $m$ with components $F_n$; $\mathbf{G}^{(m)}$ in (39) is a square matrix of dimensions $2m \times 2m$, which has the following form:

$$\mathbf{G}^{(m)} = \begin{pmatrix} \mathbf{E}^{(m)} & h\mathbf{F}^{\dagger(m)} \\ -h\mathbf{F}^{(m)} & \mathbf{E}^{(m)} \end{pmatrix}. \tag{40}$$

Here, $\mathbf{E}^{(m)}$ is the identity (unity) $m \times m$ matrix, $\mathbf{F}$ is the lower triangular Toeplitz $m \times m$ matrix of the form:

$$\mathbf{F}^{(m)} = \begin{pmatrix} F_0 & 0 & 0 & \cdots & 0 \\ F_1 & F_0 & 0 & \cdots & 0 \\ F_2 & F_1 & F_0 & \cdots & 0 \\ \vdots & \vdots & \vdots & \ddots & 0 \\ F_{m-1} & F_{m-2} & F_{m-3} & \cdots & F_0 \end{pmatrix}, \tag{41}$$

and $\mathbf{F}^{\dagger(m)}$ is the Hermitian conjugate of the matrix $\mathbf{F}^{(m)}$.

One can see that $\mathbf{F}^{(m)}$ is a Toeplitz $m \times m$ matrix and, as a result, $\mathbf{G}^{(m)}$ is also a Toeplitz matrix, which is however non-Hermitian (in contrast to that considered in [43]). In order to solve Eq. (39), it is necessary to determine the inverse of the matrix $\mathbf{G}^{(m)}$. Owing to the special structure of a Toeplitz matrix, its inversion can be obtained using fast algorithms, such as the fast algorithm proposed in [45]. After obtaining the inverse of $\mathbf{G}^{(m)}$, the $m^{\text{th}}$ element of the BNFT can be calculated using the association (38).

However, the direct utilization of the approach from [43] for the inversion of a Hermitian Toeplitz matrix implies that at each step one has to update the matrix $\mathbf{F}^{(m)}$ with one row and one column; i.e., we have a rank 1 update for $\mathbf{F}^{(m)}$ for each consecutive iteration, $m = 1, 2... N$. In turn, this means that the rank of the whole matrix $\mathbf{G}^{(m)}$ increases by 2 when one proceeds to the next iteration step, $m + 1$. Thus, one cannot apply directly the iteration scheme for the inversion of the non-Hermitian Toeplitz matrix from Zohar's method [45], as it assumes the rank 1 matrix update at each iterative step. Therefore, for the non-Hermitian case considered in our study, we generalize the approach from [43]: at every step, say number $m$, of our algorithm (which can be named the outer broadening scheme, in contrast to the inner broadening suggested in [43]) we use the modified matrix $\widetilde{\mathbf{G}}^{(m)}$ with the rank $N + m$:

$$\widetilde{\mathbf{G}}^{(m)} = \begin{pmatrix} \mathbf{E}^{(N)} & h\mathbf{F}^{\dagger(m)} \\ -h\mathbf{F}^{(m)} & \mathbf{E}^{(m)} \end{pmatrix}, \qquad (42)$$

where, as before, $\mathbf{E}^{(m)}$ is the identity matrix of the corresponding rank. One can see that this modified matrix (42) is obtained by taking the first $N + m$ rows and columns of the general rank $2N$ matrix $\mathbf{G}^{(N)}$ from Eqs. (39), (40). Then, at the $m$-th step we also redefine the modified right hand side vector (cf. Eq. (39)):

$$\widetilde{\mathbf{b}}^{(m)} = [\underbrace{0, 0, ...0}_{N}, F_0, F_1, ...F_{m-1}]^T \qquad (43)$$

The dimension of $\widetilde{\mathbf{b}}^{(m)}$ is obviously $N + m$. The inverse matrix $(\widetilde{\mathbf{G}}^{(m)})^{-1}$ at each step is now obtained by the straightforward application of the iterations used in Zohar's method. Finally, one convolves the modified inverse matrix with the modified right-hand-side vector, $(\widetilde{\mathbf{G}}^{(m)})^{-1} \cdot \widetilde{\mathbf{b}}^{(m)}$, and takes the last element of the resulting vector to get $\widetilde{v}_{(m)}^{(m)}$; note that its value is exactly the same as it would be for $v_m^{(m)}$ if one applies the inner broadening and rank 2 updating from work [43]. The desired distribution in the time domain, $q^{(m)}$, is again obtained by using Eq. (38) with $\widetilde{v}_{(m)}^{(m)}$ inserted. At the next $(m + 1)$-th step we take $N + m + 1$ rows and columns from the full matrix $\mathbf{G}^{(N)}$, Eq. (40), to obtain the next $\widetilde{\mathbf{G}}^{(m+1)}$ Toeplitz matrix, so that the rank of the iterative update for our method is just 1, which allows us to perform the next Zohar's iteration [45]. Subsequently, one composes the consecutive right hand-side vector $\widetilde{\mathbf{b}}^{(m+1)}$, convolves it with $(\widetilde{\mathbf{G}}^{(m+1)})^{-1}$ to get $\widetilde{v}_{(m+1)}^{(m+1)}$, and then recovers the next value in the time domain $q^{(m+1)}$. Repeating these operations for $m = 1, 2... N$ gives us the complete recovery of the profile $q(t)$ for the desired time-domain interval. Note that although here we consider a non-Hermitian case corresponding to the anomalous dispersion NLSE, the same idea with the rank 1 matrix updating can be applied to the inversion of the Hermitian matrix, which in the case of the GLME corresponds to the normal dispersion NLSE [24] or in a number of problems referring to the Bragg grating synthesis [48].

In order to confirm the validity of this numerical approach, we take into the following time-domain signal, whose exact BNFT is known [49]:

$$F(t) = v\alpha e^{-\alpha t} \qquad (44)$$

where $\alpha > 0$ and $-1 \leq v \leq 1$. The exact solution for the BNFT with F(t) given by Eq. (44) is:

$$q(t) = -\frac{4\alpha v \sigma (\sigma - 1)}{(\sigma - 1)^2 e^{-2\sigma \alpha t} + v^2 e^{2\sigma \alpha t}}, \qquad (45)$$

where $\sigma = \sqrt{1+v^2}$.

The numerical and analytical results for the BNFT of $F(t)$ are compared in Fig. 3. The RMSE of our numerical BNFT method, normalized by the peak value of $F(t)$, is shown in Fig. 3(c) as a function of the time resolution $dt$. A similar behavior with the FNFT data behavior has been observed before (see Fig. 2(c)), where the RMSE value also increases with the growth of the input signal power. This increase of the numerical error imposes limitations to the signal launch power in the NIS-based systems.

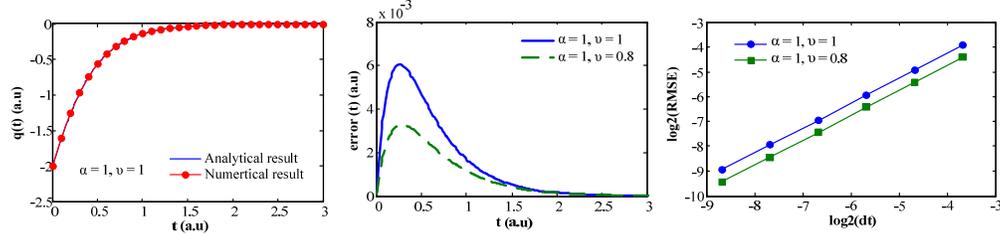

Fig. 3. Comparison between the numerical and analytical results. (a) Numerical and analytical solution for $q(t)$, $dt = 0.01$. (b) Error of numerical method as a function of t. (c) RMSE as a function of the time resolution $dt$.

## 4.3. Numerical complexity of the NIS method and of the related NFT operations

An important quantity to consider with regard to the NIS (and generally NFT-based) transmission methods is the numerical complexity (the number of floating point operations) of the NFT-based processing. This can be compared to, e.g., the popular DBP technique for the removal of nonlinear signal distortions [3]. For the latter, one reads the transmitted waveform at the receiver, inserts it as an input for the noiseless NLSE (5), swaps the sign of $z$, and then solves the NLSE in a backward direction down to the initial point $z = 0$. The numerical solution of the NSLE is usually performed by using different modifications of a Split-Step Fourier method (SSFM) [3, 10], which typically requires $\kappa M_z N \log(N)$ floating point operations. Here, the value of the numerical factor $\kappa$ depends on the order (and type) of the SSFM and it is usually of the order of 10 or more, where $N$ is the number of discretization points in the time domain, and $M_z$ the number of steps in $z$ (the distance along the fiber). $M_z$ is linearly proportional to the overall transmission length and can also depend on the pulse power when the latter is high enough and the elementary "dispersion step" in $z$ becomes comparable to or larger than a "nonlinear step"; see [10] for more details and direct references with regard to the SSFM's performance and accuracy. In realistic problems, for long-haul transmission lines one typically has $M_z \gg 1$. In turn, the NIS method involves just two separate nonlinear transforms, GLME (BNFT) at the transmitter and ZSP (FNFT) at the receiver [30]. Each of the NFTs requires $\sim N^2$ floating-point operations with the use of well-developed "traditional" methods; see [40-42] for the estimations and comparison of the PCA method used in our paper, and [43] and references therein for the Hermitian BNFT based on the Toeplitz matrix inversion applied to the Bragg gratings synthesis. Even with this estimation in mind, the numerical complexity of the NIS can already be comparable with that of DBP for sufficiently long transmission lines [29]. However, the recent advancements in numerical NFT indicate that the complexity of NIS can be potentially reduced even further than $\sim N^2$ operations. In this work, for the solution of the GLME (BNFT) we utilize the method based on the Toeplitz matrix inversions [43]. A number of works propose stable "superfast" algorithms for the Toeplitz matrix inversion [46,47], where the reported number of the floating point operations is only $\sim N \log^2(N)$ or $\sim N \log^3(N)$, so that it is generally comparable with the numerical complexity of a single FFT operation (the recent advancements in superfast Toeplitz matrix inversion methods are summarized in [47].) This means that it is already approximately of the same order as required for a single step of the SSFM used in the DBP method. Another direction in the development of the NIS approach is the increase of the precision of the BNFT, which actually limits the performance of the NIS;

see Fig. 3(c). In particular, in [43] the authors suggested the usage of higher-order integration schemes when proceeding to the matrix equations, to gain a higher accuracy for the Hermitian GLME solution while keeping the same order of the numerical complexity; this approach has yet to be generalized to the non-Hermitian (non-soliton) GLME occurring in the case considered in our paper. Potentially, there are many different variants for the BNFT methods [43,45,49] applicable to the Hermitian version of the GLME, most of which can be generalized to the case of the non-Hermitian GLME (without solitons), opening the possibility for a versatile design of the NFT-based processing schemes. For the ZSP (FNFT) (Eq. (9)), very recent studies [48, 49] suggest that the recovery of the continuous part of the nonlinear spectrum, utilized further in our study, can be achieved in only $\sim N \log^2(N)$ operations; actually, this fact demonstrates the advantage of the NIS method based exactly on the continuous part of the nonlinear spectrum, since for the solitonic part the complexity of the FNFT and BNFT can be sufficiently higher. Therefore, this complexity is again comparable with that of just a single SSPM step. The other variants of the numerical method for the FNFT [27,50,51] can also be considered for our situation. Because of the estimations given above, we believe that the NIS-based transmission methods can be highly competitive and even eventually outperform DBP in terms of numerical complexity for the digital signal processing, especially when long-haul transmission is addressed. With the application of the methods having higher accuracy, the performance characteristics of NIS can also be potentially enhanced further, and the availability of different solutions for the numerical NFT operations makes the particular design of the NIS-based transmission lines fairly flexible.

## 5. Simulation results and discussion

In this section, we present the study of the performance of high-bit-rate, high-SE NIS-based systems in comparison with systems employing DBP and having the same parameters. As the NIS method is based on Eq. (5), this scheme is specifically appropriate to the single polarization transmission. Developing an appropriate NIS scheme for the polarization division multiplexed (PDM) transmission is currently out of the scope of our paper. However, we note that the NIS ideology could be potentially applied to the birefringent fibers, where the equalization (on the average) of the nonlinear self- and cross-phase modulation coefficients in the equations for the pulse components results in the interable two-component generalization of the NLSE (so-called Manakov equation. [52]).

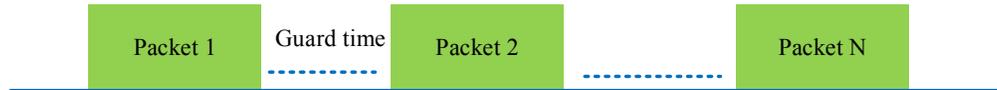

Fig. 4. Illustration of a burst mode transmission, in which neighbouring packets are separated by a guard time

As the FNFT and BNFT are performed here with the signal decaying to zero at the far ends, the proposed NIS approach is appropriate for the burst mode transmission (Fig. 4) of a multi-access network, in which neighbouring packets are separated by a guard time. The guard time duration is usually chosen longer than the channel memory, which, in our case, is the fiber chromatic dispersion induced memory. Different packet data can be sent from the same or different transmitters. In this work, for simplicity we assume that all packet data are from the same transmitter. The more general case of periodic boundary conditions will be considered elsewhere.

As mentioned before, we take into account the high-SE transmission modulation schemes, namely, the OFDM and single-carrier system with Nyquist pulse shaping (Nyquist-shaped). We designed 56-Gbaud OFDM and Nyquist-shaped NIS-based transmission systems (in burst mode) with high SE-modulation formats; namely, QPSK, 16QAM, and 64QAM. The net data rates of these systems, after removing 7% overhead due to the FEC, were 100 Gb/s, 200 Gb/s, and 300 Gb/s, respectively. The guard time duration is chosen as 20% longer than the fiber chromatic dispersion induced memory for a 2000km-link. For the OFDM NIS-based system, the IFFT size was 128, where 112 subcarriers were filled with data (with Gray-coding) while the remaining subcarriers were set to zero. The useful OFDM symbol duration was 2 ns and no cyclic prefix was used for the linear dispersion removal. After the IFFT, the time domain OFDM signal was fed into the BNFT block. For simplicity we assume that each packet data contains only one OFDM symbol. In the case of Nyquist-shaped system, each packet data contains 128 symbols. An oversampling factor of 40 was used for both OFDM and Nyquist-shaped system, resulting in a total simulation bandwidth of around 2THz. The oversampling factor of 40 was adopted to diminish the numerical errors associated with the BNFT and FNFT for achieving an accurate estimate of the performance for the NIS approach. In this work, all the DSPs are performed for the same value of the oversampling factor (i.e. 40).

The transmission link was assumed to be lossless (with ideal Raman amplification), for which the ASE noise density is expressed through the line parameters as [2]:

$$N_{ASE} = \alpha L h f_s K_T \qquad (46)$$

where $\alpha$ is the fiber loss, $L$ is the transmission distance, $hf_s$ is the photon energy, $f_s$ is the optical frequency of the Raman pump providing the distributed gain, and $K_T$ is the photon occupancy factor, which is taken to be equal to 1.13 for the Raman amplification of the fiber-optic communication systems at room temperature [2]. In our simulation, we assumed that the long-haul fiber link consisted of 80-km spans of a standard single-mode fiber (SSMF) with a loss parameter of 0.2 dB/km, nonlinearity coefficient of 1.22 $W^{-1}km^{-1}$, and dispersion of 16 ps/nm/km. To demonstrate a more realistic transmission condition we used a photon occupancy factor of 4 in our simulation, and the ASE noise was added after each fiber span. At the receiver, after coherent detection, the FNFT was performed by employing the PCA method (see Subsection 4.1) to obtain the nonlinear spectrum (left reflection coefficient, see Eq.(15)). The single-tap dispersion removal was then performed to recover the nonlinear spectrum of the transmitted signal, which was followed by the IFFT operation and then the feeding of the outcome into the standard decoder. In this work we assume having a perfect timing synchronization, the transmitter laser and the local oscillator are noiseless and no frequency offset is considered. For simplicity, we consider only the transmission of a single packet. Monte-Carlo simulation is then performed to estimate the system performance using the error vector magnitude (EVM) and direct error counting method. For convenience, the estimated system BER is then converted into Q-factor [54] for further discussion

*5.1. NIS performance without the ASE noise*

Firstly, we consider the case when the ASE noise is ignored. The linear spectra of OFDM signals before and after the BNFT are shown in the Fig. 5. It can be seen that after the BNFT, the linear spectrum of the OFDM signal does not broaden significantly, indicating that the NIS method combined with the OFDM can be effectively applied for a WDM transmission or even multiplexed into superchannels. However, the performance of the NIS scheme for the WDM transmission is beyond of the scope of this paper, and here we consider only single channel transmission. In Fig. 6, we compare the performance of the 100 Gb/s OFDM and Nyquist systems with and without the NIS method for fiber nonlinearity compensation. Herein, the fiber nonlinearity is the only channel's impairment. The modulation format used there was the QPSK. The performance indicator is the Q-factor, which is calculated through the EVM [53,54]. In Fig. 6, almost no mismatch is observed between the back-to-back performance (without ASE noise) and the performance after 2000 km of SSFM for NIS-based systems. This result confirms that the NIS-based approach can perfectly compensate for the deterministic impairment due to the fiber nonlinearity, using just a single-tap linear dispersion removal for the nonlinear spectrum at the receiver. This result demonstrates the potential of the NIS method as a novel alternative approach for compensating the fiber nonlinearity impairments in optical communication. However, one can notice that the back-to-back performance of NIS-based systems deteriorates when the input signal power increases. This phenomenon can be explained by the fact that the numerical error of both FNFT and BNFTs grows with the increase of input signal power, as shown in Figs. 2 and Fig. 3. Consequently, the accuracy of the FNFT and BNFT operations imposes a fundamental limit on the NIS-based transmission systems. The reduction of the relative error of the FNFT and BNFT is achieved by increasing the sampling rate. However, this approach cannot be ultimately effective because of the limitation of the time sampling resolution in the current ADC/DAC technology. A more practical but more challenging approach would be to develop more accurate and efficient algorithms for performing the backward and forward NFTs. As a result, a lot of efforts have yet to be applied in this direction to make the NIS method's performance more efficient, keeping in mind the constrains in the time sampling resolution.

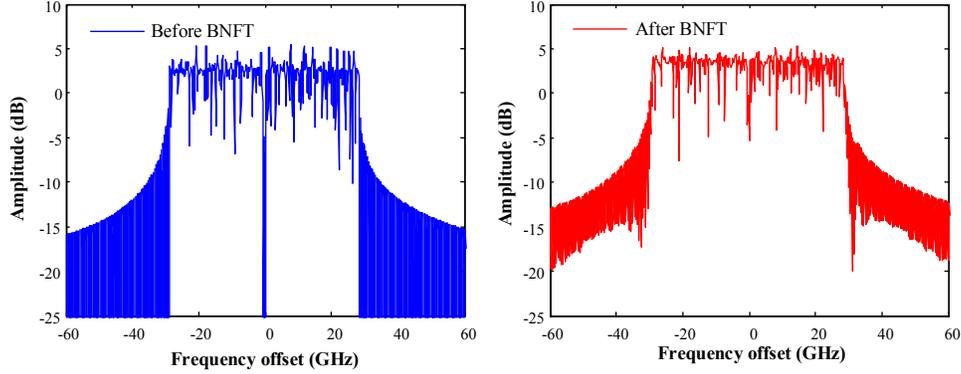

Fig.5. Linear spectra of OFDM signals before and after BNFT, the launch power is 0dBm.

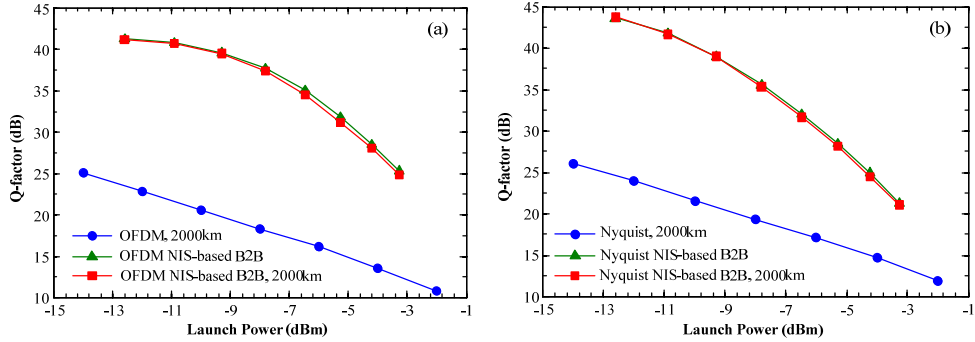

Fig.6. Q-factor as a function of the launch power for the (a) OFDM and (b) Nyquist-shaped NIS-based systems without the ASE noise.

*5.2. Performance comparison of NIS versus DBP in the presence of ASE noise*

In this subsection, we compare the performance of the OFDM and Nyquist-shaped systems with the use of the NIS and DBP methods for fiber nonlinearity compensation. For the implementation of DBP, the received signal is first filtered with an $8^{th}$ order low-pass filter having a bandwidth of 40 GHz. Subsequently, the optical field is reconstructed and the signal is back-propagated with a different number of steps per single span, indicating the numerical complexity of the corresponding DBP realization.

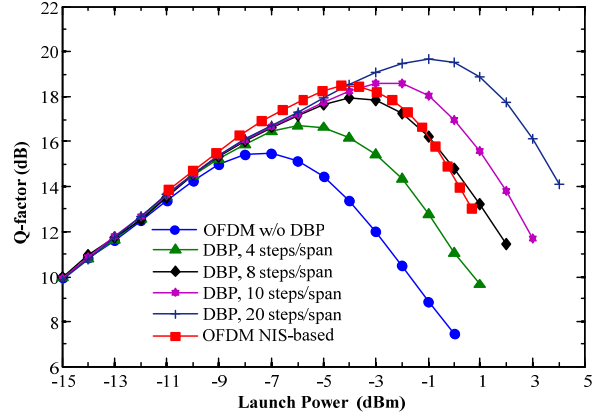

Fig. 7. Performance comparison of the 100-Gb/s QPSK OFDM systems with the NIS vs. the DBP methods for fiber nonlinearity compensation. The receiver filter bandwidth used was 40 GHz, the distance is 2000km.

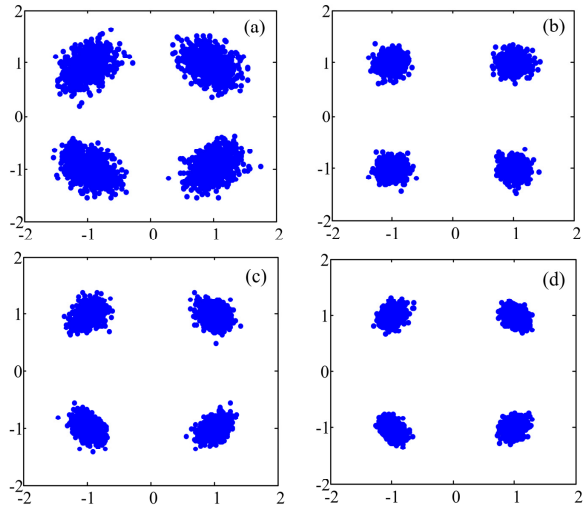

Fig. 8. Constellation diagrams at the optimum launch powers of the 100-Gb/s QPSK OFDM systems with and without the NIS and DBP methods for fiber compensation; (a) without NIS and DBP, (b) with the NIS method, (c) DBP with 10 steps/span, (d) DBP with 20 steps/span

In Fig. 7, we compare the Q-factors of OFDM systems with NIS and DBP. One can see that the OFDM NIS-based system offers over 3.5 dB advantage over the traditional OFDM system, confirming the effectiveness of the proposed approach for fiber nonlinearity compensation. This performance improvement is comparable with that of DBP with 10 steps per span. The launch power in the NIS-based system is limited to -4 dBm (the optimum launch power), which we believe is mainly due to the numerical errors of the backward and forward NFT at the transmitter and receiver, respectively. This argument is also supported by the fact that the Q-factor of the OFDM NIS-based system decreases faster than that of the DBP for the high-input powers. The constellation diagrams of the OFDM systems with NIS and DBP recorded at the corresponding optimum launch powers are shown in Fig. 8. It should be noted that the received constellation diagram in system with 10 steps per span DBP clearly indicates the presence of residual nonlinear phase noise. On the other hand, the constellation diagram in system with NIS approach looks "rounded", indicating that nonlinear phase noise has been fully compensated. The main residual impairment here is the numerical error of the FNFT and BNFT. For the QPSK modulation format, the Q-factor was calculated through the EVM [53].

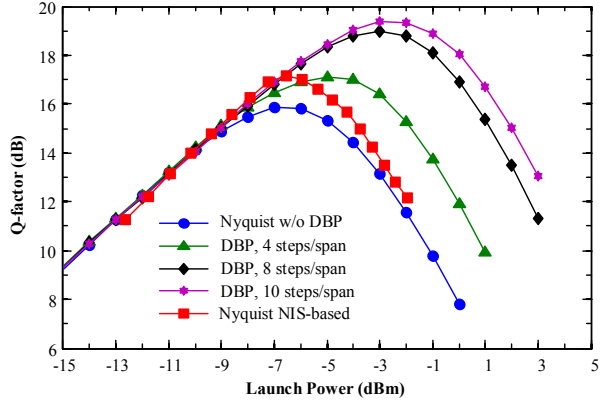

Fig. 9. Performance comparison of the 100-Gb/s QPSK Nyquist-shaped systems with the NIS and DBP methods for fiber nonlinearity compensation. The receiver filter bandwidth is 40 GHz, the distance is 2000km

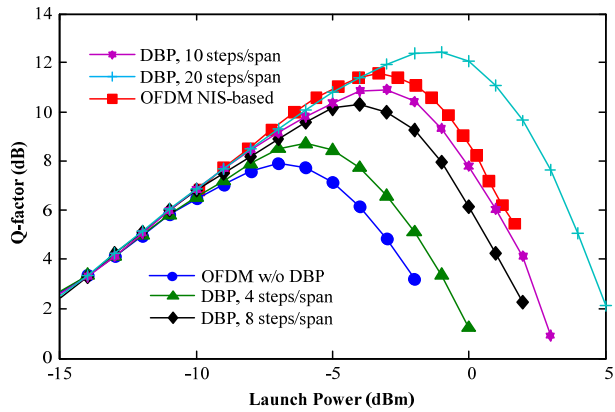

Fig. 10. Performance comparison of the 200-Gb/s 16QAM OFDM systems with the NIS and DBP methods for fiber nonlinearity compensation. The receiver filter bandwidth was 40 GHz, the distance is 2000kmand DBP, (b) DBP with 20 steps/span, (c) with the NIS method.

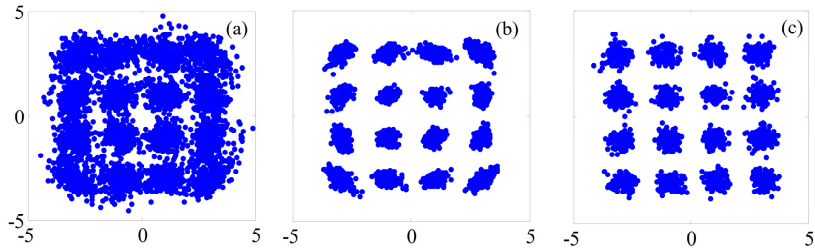

Fig. 11. Constellation diagrams at the optimum launch powers of the 200-Gb/s 16QAM OFDM systems with and without the NIS and DBP methods for fiber compensation. (a) Without NIS

In Fig. 9, the performance of the 100 Gb/s Nyquist-shaped systems with and without the NIS and DBP techniques is compared. It is evident that for the NIS method a performance improvement of about 1.5 dB can be achieved. This improvement is comparable with what can be realized with the DBP approach using 4 steps per span. In our simulations, we observed that the OFDM and Nyquist-shaped systems have similar performance, which agrees well with some current numerical and experimental comparisons of the OFDM and Nyquist-shaped systems [55,56]. However, when combining it with the NIS method, OFDM displays a much better performance. In the considered system configuration, the optimum Q-factor of the OFDM NIS-based system is around 19 dB, while for the Nyquist-shaped NIS-based system the optimum Q-factor is approximately 17.5 dB. We also notice that the OFDM NIS-based system has a better performance than the Nyquist-shaped NIS-based one in the back-to-back regime. This indicates that the numerical error associated with the NFT transformations for the same power is smaller for the OFDM than for the Nyquist-shaped signal. This is due to the fact that the OFDM signal has a smaller $L1$-norm in comparison with that of the Nyquist-shaped signal having the same power. Another advantage of the OFDM over the Nyquist-shaped signal for the NIS-based system is that after the dispersion unrolling, the IFFT block is not required for OFDM NIS-based systems because in the OFDM signal the information is encoded in the frequency domain, and hence the conversion to the time domain is not necessary. As a result, we conclude that the OFDM modulation is a more suitable modulation format for NIS-based systems in comparison to the Nyquist-shaped signal. Consequently, for higher order modulation formats, such as 16QAM or 64QAM, we employ further only the OFDM modulation of the nonlinear spectrum in combination with the NIS method.

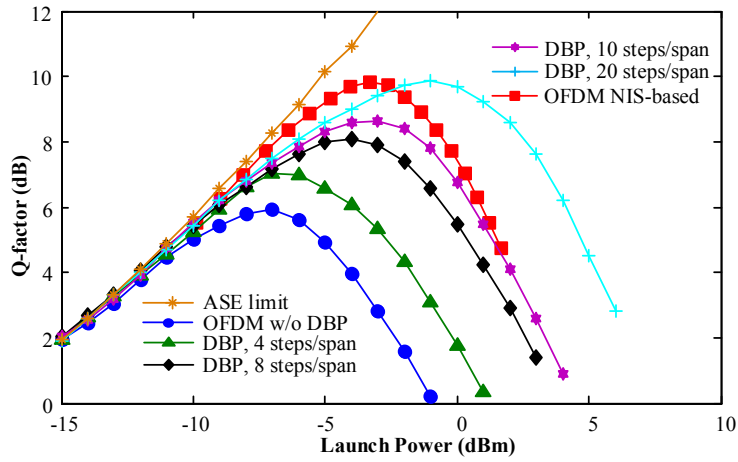

Fig. 12. Performance comparison of the 300-Gb/s 64QAM OFDM systems with the NIS and DBP methods for fiber nonlinearity compensation. The receiver filter bandwidth was 40 GHz, the distance is 800km

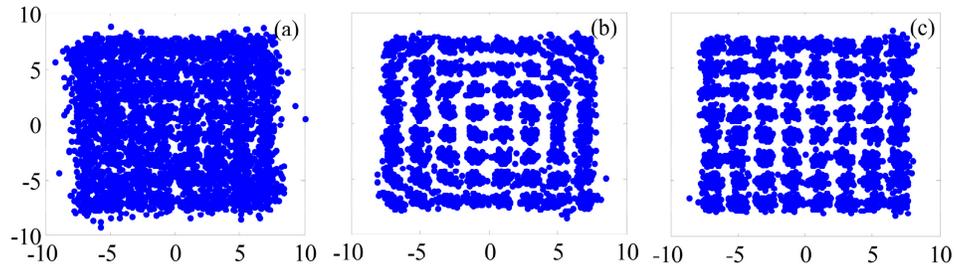

Fig. 13. Constellation diagrams at the optimum launch powers of the 300-Gb/s 64QAM OFDM systems with and without NIS and DBP methods for fiber compensation. (a) Without NIS and DBP, (b) DBP with 20 steps/span, (c) with the NIS method.

When combining it with a higher modulation format, such as 16QAM, the OFDM NIS-based approach offers nearly 4 dB advantage over the traditional OFDM scheme; see Fig. 10. The transmission bit rate in this case was increased to 200 Gb/s; the performance indicator Q-factor was also calculated using the EVM [53]. It can be seen that for the 16QAM modulation format, the OFDM NIS-based system outperforms the DBP with 10 steps per span. The optimum constellation diagrams for the conventional OFDM system, OFDM NIS-based, and OFDM with 20 steps per span DBP are shown in Fig. 11. This Fig. shows that fairly clear constellation diagrams can be achieved with the NIS method.

The simulation results for the 300-Gb/s 64QAM OFDM NIS-based system are compared in Figs. 12 and 13 with the conventional OFDM and OFDM with DBP. In these simulations, the performance indicator Q-factor was calculated through direct error counting. It can be seen that for such high-order modulation format, the OFDM NIS-based system displays almost the same performance as that of the DBP with 20 steps per span. The performance improvement in comparison with the conventional OFDM system is about 4.5 dB, which is larger than the values achieved for the QPSK and 16QAM modulation formats. This result indicates that a greater performance advantage of the OFDM NIS-based system over the traditional approaches can be reached for higher-order modulation formats, and shows the considerable benefit of the NIS method for fiber nonlinearity compensation for high-SE transmission systems. In Fig. 12 we also present the curve indicating the ASE transmission limit: For calculating it, we completely removed the nonlinearity. It can be seen that the curve for the NIS-based transmission generally goes above those for the DBP in the noise-dominated region. However, it does not intersect the limiting line. This behavior reveals that the NIS-based transmission is less sensitive to the noise-induced corruption than the DBP, and the refinements of the NFT processing techniques can improve the NIS performance even further.

## 6. Conclusion

In this work, we have shown that the NIS method can be successfully combined with high-SE transmission techniques (e.g., OFDM, Nyquist-shaped) and advanced modulation formats, such as QPSK, 16QAM, and 64QAM. This novel transmission scheme suggests the encoding of the information onto the continuous part of the nonlinear spectrum and requires only single-tap equalization at the receiver to compensate for all the deterministic fiber nonlinearity impairments accumulated along the fiber link. Generally, the NIS concept considered in our work can be further extended to other optical systems described by different integrable continuous equations, like, e.g., Manakov equation, governing the averaged pulse evolution in randomly birefrigent fibers. Our simulations confirmed the effectiveness of the NIS scheme and showed that an improvement of 4.5 dB can be achieved, which is comparable with the DBP compensation method employing multi-steps per span. We would like to emphasize that with the utilization of increasingly higher order modulations (16- and 64-QAM), the results for the performance of the NIS-based OFDM system became comparable with the performance of increasingly higher-order DBP compensation (i.e., with progressively more steps per span). This fact indicates that the NIS method's efficiency can become strongly competitive with that of the DBP methods for the high-SE formats, especially when one employs more accurate advanced techniques for the numerical NFT processing. We have also compared the performance of the OFDM and Nyquist-shaped modulation formats combined with the NIS method and demonstrated that OFDM is potentially a more suitable modulation for such systems.

**Acknowledgment**

The support under the UK EPSRC programme Grant UNLOC (EP/J017582/1) is gratefully acknowledged. The authors are grateful to Leonid Frumin and Andrew Ellis for helpful comments and discussions.